# Superconductivity at 36 K in β-Fe$_{1.01}$Se with the Compression of the Interlayer Separation Under Pressure

S. Medvedev[1,2], T. M. McQueen[3], I. Trojan[2,4], T. Palasyuk[2,5], M. I. Eremets[2], R. J. Cava[3], S. Naghavi[1], F. Casper[1], V. Ksenofontov[1], G. Wortmann[6] and C. Felser[1]

[1]Institut für Anorganische und Analytische Chemie, 55099 Mainz, Germany
[2]Max-Planck-Institute for Chemistry, 55128 Mainz, Germany, Germany
[3]Department of Chemistry, Princeton University, Princeton NJ 08544, USA
[4]I A.V. Shubnikov Institute of Crystallography, 119333 Moscow, Russia
[5]Institute of Physical Chemistry PAS, Warsaw 01-224, Poland
[6]Department Physik, Universität Paderborn, 33095 Paderborn, Germany

**In this letter, we report that the superconducting transition temperature in β-Fe$_{1.01}$Se increases from 8.5 to 36.7 K under an applied pressure of 8.9 GPa. It then decreases at higher pressures. A dramatic change in volume is observed at the same time T$_c$ rises, due to a collapse of the separation between the Fe$_2$Se$_2$ layers. A clear transition to a linear resistivity normal state is seen on cooling at all pressures. No static magnetic ordering is observed for the whole p-T phase diagram. We also report that at higher pressures (starting around 7 GPa and completed at 38 GPa), Fe$_{1.01}$Se transforms to a hexagonal NiAs-type structure and displays non-magnetic, insulating behavior. The inclusion of electron correlations in band structure calculations is necessary to describe this behavior, signifying that such correlations are important in this chemical system. Our results strongly support unconventional superconductivity in β-Fe$_{1.01}$Se.**

**Introduction**

Superconductivity has recently been discovered in iron arsenides [1-6], with superconducting transition temperatures ($T_c$'s) as high as 55 K [6]. The superconductivity in this class of materials is unexpected because most Fe-based compounds display strong magnetic behavior. The iron arsenides share a number of general features with the high-temperature superconducting cuprates, including high $T_c$'s, proximity to a magnetically ordered state, and a linear temperature dependence of the resistivity [7,8]. These properties suggest that the superconductivity in the iron arsenides is unconventional, with electron pairing possibly mediated by magnetic interactions. More recently, superconductivity has been reported at 8.5 K [9], increasing up to 27 K under pressure [10], in the structurally related material iron selenide [11,12]. Iron selenide and the iron arsenides share a common structural motif, containing $Fe_2X_2$ (X=As, Se) layers of edge-shared $FeX_4$ tetrahedra, suggesting that the mechanism of superconductivity may be the same in both families in spite of the low $T_c$ for the selenide.

Here we show that superconducting $\beta$-$Fe_{1.01}$Se [11] exhibits a transition to an anomalous linear temperature dependence of the resistivity in the normal state at temperatures above $T_c$, stable over a wide range of applied pressures. This linear dependence in the resistivity in $\beta$-$Fe_{1.01}$Se is preserved as the superconducting temperature rises to 36 K at 8.9 GPa. Taken in the context of recent pressure dependent NMR measurements [13], this suggests that the anomalous normal state resistivity is related to both the spin fluctuations and superconductivity in $\beta$-$Fe_{1.01}$Se. We also show that unlike the iron arsenides, no static magnetic ordering is observed in $\beta$-$Fe_{1.01}$Se at pressures up to 31 GPa. This observation is consistent with the electronic structure calculations of Lee et al., which conclude that a non magnetic ground state is most stable for iron selenide [14]. This suggests that if there is a magnetically ordered state to be found for this material, making it strictly analogous to the iron arsenide superconductors, such an ordered state is far away in temperature, pressure, and composition from what is presently known. At the highest pressures studied, $Fe_{1.01}Se$ transforms to a hexagonal close packed NiAs-type structure that exhibits semiconducting behavior when local density approximation band structure calculations indicate that metallic behavior should be observed. These results suggest that strong electron correlations (such as the short-lived spin correlations recently observed in β-$Fe_{1.01}Se$ [13]) are necessary for the observation of high transition temperatures in iron-based superconductors.

**Results**

In Fig. 1 (a), the room temperature X-ray diffraction patterns are shown for β-$Fe_{1.01}Se$ for various pressures. The corresponding structures are displayed in Fig 1 (b,c). At low pressures, the room temperature structure of β-$Fe_{1.01}Se$ is tetragonal (Fig. 1 (c)). At 1.5 GPa there is a ~10% reduction in the unit cell volume and a drop in the c/a ratio of the tetragonal cell parameters from 1.464 to 1.414 (3.4%) (see data points in Fig. 1 (d)). This correlates with a strong increase in $T_c$ (more than 10 K at 0.8 GPa). These two structural effects result from a collapse of the van-der-Waals-bonded region that separates the $Fe_2Se_2$ planes under pressure (blue arrows, Fig. 1(c)), and implies, in conjunction with the Mössbauer measurements (see below), a dramatically reduced interlayer separation as $T_c$ rises. By 12 GPa a part of the sample has transformed into the hexagonal NiAs-type structure (Fig. 1 (b)). This is expected as the volume of the hexagonal phase is smaller than the volume of the tetragonal phase. The very wide two-phase hexagonal + tetragonal phase region, which extends from pressures of about 7 GPa (see below) to 35 GPa on increasing pressure, is characteristic of a first order transition. By 38 GPa, only the hexagonal phase is present. Releasing the pressure back to ambient results in the full recovery of the superconducting β-Fe$_{1.01}$Se tetragonal phase, indicating, given the extreme sensitivity of the superconductivity of β-Fe$_{1.01}$Se to stoichiometry [11], that the hexagonal high pressure phase is also very close to stoichiometric. An analogous reversible transition from the tetragonal phase into the hexagonal phase was observed in thin FeSe-films [15].

The transport and magnetic properties of FeSe are very sensitive to applied pressure. To illustrate this, representative temperature-dependent resistivity curves and the Mössbauer spectra obtained at different pressures are shown in Figs. 2 (a) and (b). The transition into the superconducting state is observed to increase from 8 K to 36.7 K on applied pressure, before decreasing again (see Fig. 2 (c)). At 0.8 GPa, the onset of the superconducting transition $T_c$ is 18.6 K, with a sharp drop to zero resistance. With increasing pressure the superconducting transition becomes broader, which is also found for the superconducting transition of the iron arsenides under pressure [16]. The initial pressure coefficient of $T_c$ between ambient pressure and 0.8 GPa is around 12.6(2) K/GPa and decreases for increasing pressures (between ambient pressure and 8.9 GPa, the pressure coefficient is 3.2(1) K/GPa). Passing the superconducting dome at pressures higher than 8.9 GPa (see Fig. 2 (c)) the resistivity curves become more complicated. The decrease of the superconducting $T_c$ once past the maximum is more moderate, approximately -1.7(2) K/GPa. At 15 GPa (Fig. 2 (c)) the superconducting transition temperature has decreased to 25 K, but the resistance does not drop to zero below the superconducting transition temperature. The residual resistance below $T_c$ is due to the occurrence of larger amounts of the semiconducting hexagonal phase at this pressure.

A change in the slope in the temperature dependence of the resistivity is observed at all pressures in the stability region of the superconducting phase. This transition occurs at temperatures between ~90 K and 60 K, depending on the pressure, and decreases in temperature with increasing pressure (Fig. 2 (c)). The transition to linear resistivity is very sharp in much of the pressure range, but undergoes a change in concavity making precise assignment of the transition temperature at 0.8 GPa difficult. At ambient pressure this change in the resistance, from linear behavior below to normal conductivity above the transition, is accompanied by the change of the structure from tetragonal to orthorhombic symmetry [11,12,17]. Although this may be the case under pressure as well, we currently have no information on this aspect of the pressure dependent behavior of this system. The resistance shows *T*-linear behavior R = $A + BT^n$, with *n* = 1.0 ± 0.03 in this temperature regime between the onset of this phase and the superconducting transition temperature, e.g. 90 K at atmospheric pressure. The change in the slope of the temperature dependent resistivity is largest and sharpest at intermediate pressures (e.g. near 5.1 GPa, Fig 2(c)). A *T*-linear resistivity is frequently taken as a fingerprint of a "non-canonical Fermi liquid system", as known for high $T_c$ cuprates [18] or in heavy Fermion metals [19,20]. This unusual linear temperature dependency of ρ(T) was also recently observed in the iron arsenides [21]. The temperature regime where we observe the linear resistivity is correlated to the region where NMR has detected the presence of increasing spin fluctuations in β-$Fe_{1.01}$Se [13]. In contrast to our findings for β-$Fe_{1.01}$Se, for $LaO_{1-x}F_xFeAs$ and $Ba_{1-x}K_xFe_2As_2$ the linear resistivity behavior and the orthorhombic structural distortion are intimately correlated to the presence of a long range ordered spin density wave which sets in at or slightly below the temperature of the structural phase transition [3,22,23].

To characterize the magnetic properties of the different phases of FeSe appearing in the pressure experiments, $^{57}$Fe-Mössbauer spectroscopy was applied as a sensitive local probe. Representative $^{57}$Fe-spectra are shown in Fig. 2 (b). For the pressures up to 5.8 GPa, the spectra are well described by a single quadrupole doublet with hyperfine parameters similar to those found for the same sample at ambient pressure [11]. The quadrupole splitting, $\Delta E_Q = 0.28(2)$ mms$^{-1}$, and isomer shift, $\delta = 0.44(1)$ mms$^{-1}$ at 0.2 GPa can be attributed to formally divalent Fe ions in a distorted tetrahedral surrounding of Se neighbors with strong covalency in the Fe-Se bonds. At 7.2 GPa a new phase appears in the spectrum with a relative intensity of ca. 20%, a distinct different isomer shift, $\delta = 0.84$ mms$^{-1}$, and a quadrupole splitting of $\Delta E_Q = 0.21(2)$ mms$^{-1}$. These values of hyperfine parameters are characteristic of formally divalent Fe in more symmetric surrounding with less covalency and can be unambiguously attributed to the hexagonal phase of FeSe [24], present in the XRD pattern (Fig. 1). The relatively small variation the Mössbauer parameter with pressure indicates only a modest variation of the local surroundings of the Fe ions especially in the tetragonal phase, reflected in the almost constant values of the quadrupole splittings in the whole pressure range. The local volume around the Fe ions in both the tetragonal and hexagonal phases, as defined by the Fe-Se and Fe-Fe distances, is changed much less under pressure than the molar volumes, especially at pressures up to 12 GPa. (see Fig. 1(c,d) and ref. 12). This supports the conclusion that the strong drop in volume and c/a ratio with pressure in the tetragonal phase is due to the collapse of the Se-Se van-der-Waals layer between the Fe$_2$Se$_2$ planes. A detailed report on the Mössbauer characterization of these phases, including an anomalous decrease of the Mössbauer-Lamb factor (i.e. the Debye-Waller factor) at pressures up to 7.2 GPa, will be presented in a forthcoming publication [25]. The spectra taken at 4.2 K at all pressures provide important information about the presence or absence of static magnetic ordering in these phases. Inspection of the inset in Fig. 2 (b) at 14.4 GPa indicates no magnetic hyperfine splitting in either subspectrum. This proves that the ground states in both the tetragonal and hexagonal phases of FeSe are nonmagnetic.

At pressures higher than 29 GPa, the bulk FeSe sample is semiconducting (Fig. 3(a)), indicating that hexagonal FeSe is non-metallic. Mössbauer spectra measured at 19.7 GPa (the sample contains 40% hexagonal phase), show that this form is also non-magnetic (Fig. 3(b)). The anomalous resistivity behavior vanishes with the transformation of the sample into the hexagonal phase, and our measurements indicate that this phase is not superconducting down to 0.6 K. To better understand the semiconducting, nonmagnetic character of the hexagonal phase, we have also performed electronic structure calculations. The hexagonal phase adopts the NiAs structure, and a full geometry optimization was performed for the calculations. The calculations, based only on DFT (density functional theory) within the LSDA (local spin density approximation), lead to the prediction of a metallic ground state (Fig. 3 (c) upper panel), which our observations show is incorrect. The hexagonal phase with the NiAs-structure type becomes semiconducting in the calculations only by the inclusion of correlations, using either Coulomb correlations via the inclusion of an on-site repulsion, $U \sim 4$ eV (Fig. 3(c), middle panel), or an exact exchange formalism via a hybrid functional (Fig. 3(c), lower panel). The experimentally observed nonmagnetic semiconducting character for high pressure hexagonal FeSe is therefore seen to be only consistent with an electronic system that displays substantial electronic correlations. For the tetragonal phase, our electronic structure calculations based on DFT using the LSDA are in agreement with the results of other groups [14], and inclusion of a hybrid functional does not change the calculated electronic structure of the metallic phase significantly. However, DFT+LSDA correctly reproduces the gross behavioral features of many metallic systems, even those in which electron correlations are important, such as the cuprates [26, 27]. Thus the fact that the inclusion of correlations is necessary to describe the hexagonal phase indirectly suggests that correlations play a role in tetragonal iron selenide. This is consistent with recent NMR results, which show the presence of short-lived spin correlations in tetragonal FeSe [13].

**Conclusion**

The electronic pressure-temperature (p-T) phase diagram for FeSe is shown in Fig. 4. At a first view the p-T phase diagram looks very similar to that of the iron arsenides. The high temperature superconductivity in β-Fe$_{1.01}$Se is intimately correlated with the observation of anomalous transport properties in the normal state, which we tentatively attribute to the presence of the spin fluctuations seen in NMR experiments [13]. High temperature superconductivity is found only in the region of the phase diagram where the resistivity shows the pronounced anomalous resistivity above T$_c$. Unlike the case of the arsenides however [3,22,23,28,29], the phase diagram does not have any region where a spin density wave or static magnetism is observed, and the increasing T$_c$ under pressure cannot in any way be attributed to the concurrent suppression of a magnetically ordered phase. The increase in T$_c$, is, however, associated with a dramatic decrease in volume due to the collapse of the space between the Fe$_2$Se$_2$ planes. Following the collapse of the interlayer spacing in the tetragonal β-FeSe phase, a first order transition into close-packed, semiconducting non-magnetic hexagonal phase is observed. This semiconducting phase cannot be described within the LSDA; correlations have to be taken into account. Whether these correlations are present in the electronically more complex tetragonal β-FeSe phase remains to be determined, but the data presented here strongly suggest that the superconductivity in tetragonal iron selenide is unconventional.

**Acknowledgements**

The work at Mainz was funded by the DFG in the Collaborative Research Center Condensed Matter Systems with Variable Many-Body Interactions (TRR 49). The work at Princeton was supported primarily by the U.S. Department of Energy, Division of Basic Energy Sciences, Grant No. DE-FG02-98ER45706. T.M.M. gratefully acknowledges support of the National Science Foundation Graduate Research Foundation program. I.T. was partly supported by the Russian Research Foundation under the grant No. 08-02-00897a and by the Presidium of RAS under the grant No.27-4.1.10. We are grateful to V. Prakapenka for X-ray diffraction measurements at GeoSoilEnviroCARS (sector 13, APS) at Argonne National Laboratory.

**References:**

1. Kamihara, Y. J., Watanabe, T., Hirano, M. & Hosono, H. Iron-based layered superconductor La[$O_{1-x}F_x$]FeAs with $T_C$ = 26 K. J. Am. Chem. Soc. 130, 3296-3297 (2008).
2. Chen, X. H., et al. Superconductivity at 43 K in $SmFeAsO_{1-x}F_x$. Nature 453, 761-762 (2008).
3. Zhao J. et al. Structural and magnetic phase diagram of $CeFeAsO_{1-x}F_x$ and its relation to high-temperature superconductivity. Nature Materials 7, 953 - 959 (2008).
4. Rotter, M., Tegel, M. & Johrendt, D. Superconductivity at 38 K in the iron arsenide $Ba_{1-x}K_xFe_2As_2$. Phys. Rev. Lett. 101, 107006 (2008).
5. Pitcher M. J., Parker D. R., Adamson P., Herkelrath S. J. C., Boothroyd A. T., Ibberson R. M., Brunellid M. and Clarke S. J., Structure and superconductivity of LiFeAs, Chem. Commun., 5918-5920 (2008).
6. Ren, Z. A. et al. Superconductivity at 55 K in iron-based F-doped layered quaternary compound Sm[$O_{1-x}F_x$]FeAs. Chin. Phys. Lett. 25, 2215-2216 (2008).
7. H. Takagi et al., Systematic evolution of temperature-dependent resistivity in $La_{2-x}Sr_xCuO_4$. Phys. Rev. Lett. 69, 2975 (1992).
8. Ando Y. et al., Electronic Phase Diagram of High-Tc Cuprate Superconductors from a Mapping of the In-Plane Resistivity Curvature. Phys. Rev. Lett. 93, 267001 (2004).
9. Hsu F. C., et al., Superconductivity in the PbO-type Structure alpha-FeSe. Proceedings of the National Academy of Sciences 105, 14262 (2008).
10. Mizuguchi Y., Tomioka F., Tsuda S., Yamaguchi T. and Takano Y., Superconductivity at 27 K in tetragonal FeSe under high pressure. Appl. Phys. Lett. 93, 152505 (2008).
11. McQueen T. M., Huang Q., Ksenofontov V., Felser C., Xu Q., Zandbergen H., Hor Y. S., Allred J., Williams A. J., Qu D., Checkelsky J., Ong N. P. and Cava R. J., Extreme Sensitivity of Superconductivity to Stoichiometry in FeSe ($Fe_{1+\delta}Se$), Phys. Rev. B 79, 014522 (2009).
12. Millican J. N., Phelan D., Thomas E. L., Leao J. B. and Carpenter E., Pressure-Induced Effects on the Structure of the FeSe Superconductor, arXiv:0902.0971 (unpublished)
13. Imai T., Ahilan K., Ning F. L., McQueen T. M., and Cava R. J., Why Does Undoped FeSe Become A High Tc Superconductor Under Pressure? arXiv:0902.3832 (unpublished)
14. Lee K.-W., Pardo V. and Pickett W. E., Magnetism Driven by Anion Vacancies in Superconducting $\alpha FeSe_{1-x}$ Phys. Rev. B 78, 174502 (2008).
15. Srivastava M. M. and Srivastava O. N., Studies of structural transformations and electrical behaviour of FeSe films, Thin Solid Films 29, 275 (1975).
16. Takahashi, H. et al. Superconductivity at 43 K in an iron-based layered compound $LaO_{1-x}F_xFeAs$. Nature 453, 376-378 (2008).
17. McQueen T. M., Williams A. J., Klimczuk T., Casper F., Ksenofontov V., Felser C., P. Stephens W., and Cava R. J.. Unpublished.
18. Orenstein J. and Millis A. J., Advances in the physics of high-temperature superconductivity, Science 288, 468 (2000).
19. Gegenwart P., Si Q., and Steglich F., Quantum criticality in heavy-fermion metals, Nat. Phys. 4, 186 (2008).

20  Mathur N.D., Grosche F. M., Julian S. R., Walker I. R., Freye D. M., Haselwimmer R. K. W. & Lonzarich G. G., Magnetically mediated superconductivity in heavy fermion compounds Nature 394, 39-43 (1998).
21  Hess C., et al., The intrinsic electronic phase diagram of iron-pnictide superconductors. arXiv:0811.1601 (unpublished)
22  Luetkens H. et al., The electronic phase diagram of the $LaO_{1-x}F_xFeAs$ superconductor. Nature Materials published online 2009 DOI: 10.1038/NMAT2396
23  Cruz C., et al., Magnetic order close to superconductivity in the iron-based layered $LaO_{1-x}F_xFeAs$ systems. Nature 453, 899-902 (2008)
24  Reddy K. V. and Chetty S. C., Phys. Stat. Solidi (a) 32 (1975) 585.
25  Ksenofontov V., McQueen T. M., Medvedev S., Trojan I., Palasyuk T., Eremets M., Cava R. J., Felser C. and Wortmann G.. Unpublished.
26  Czyzyk M. T., Sawatzky G.A., Local-density functional and on-site correlations: The electronic structure of $La_2CuO_4$ and $LaCuO_3$, Phys. Rev. B 49, 14211 (1994)
27  Gyorffy B. L., Szotek Z., Temmerman W. M., Andersen O. K., and Jepsen O., Quasiparticle spectra of high-temperature superconductors, Phys. Rev. B 58, 1025 (1998).
28  Drew A.J. et al. Coexistence of static magnetism and superconductivity in $SmFeAsO_{1-x}F_x$ as revealed by muon spin rotation. Nature Materials (2009) online DOI:10.1038/NMAT2397.
29  Lynn J. W. and Dai P., Neutron Studies of the Iron-based Family of High $T_C$ Magnetic Superconductors, arXiv:0902.0091 (unpublished).
30  Eremets, M. I., V. V. Struzhkin, et al., Superconductivity in boron, Science 293 272-274 (2001).
31  Blaha P., Schwarz K., Madsen G. K. H., Kvasnicka D., and Luitz J., WIEN2K, (Karlheinz Schwarz, Technische Universität Wien, Wien Austria, 2001).
32  Dovesi R. et al., CRYSTAL 2006 User's Manual (University of Torino, Torino, 2006).

**Methods**

High quality samples of a single phase were formed from a composition close to $Fe_{1.01}Se$. Details of the sample preparation were published elsewhere [11]. The samples were characterized by laboratory XRD [11]; no impurity phases were detected in all samples investigated here. To study the phase diagram, external pressure as a well defined tuning parameter was applied up to 40 GPa. The unique combination of different methods enabled us to investigate the structure by XRD, the electronic properties by resistivity measurements [30] and the magnetic properties by Mössbauer spectroscopy under pressure for the same well defined sample between room temperature and 4.2 K.

$^{57}Fe$ Mössbauer spectra were recorded using a constant-acceleration spectrometer, a helium bath cryostat and a $^{57}Co(Rh)$ Mössbauer point source with an active spot diameter of 0.5 mm. Powder FeSe samples enriched with $^{57}Fe$ (20%) were measured in a diamond-anvil pressure cell (DAC) with silicon oil as the pressure transmitting medium, enabling quasihydrostatic pressure measurements up to 40 GPa in the temperature range 4.2-300 K. Isomer shift values are quoted relative to α-Fe at 293 K.

A diamond anvil cell (DAC) was used for electrical resistance measurements under high pressures up to 29 GPa. We used a cBN/epoxy mixture for insulating gaskets and platinum foil for electrical leads[i]. The diameter of flat working surface of diamond anvil was 0.5 mm and the diameter of the hole in the gasket was 0.07 mm. The hole was filled with powder sample. We measured resistance with DC current source in van der Pauw geometry of electrodes. Pressure was measured at low and room temperatures by the Ruby scale from a small chips scattered across the sample[ii]. The pressure distribution was ≤0.05 GPa across the sample. Temperature was measured with a calibrated Si-diode attached to the DAC with accuracy within 0.1 K. The diffraction patterns were accumulated with a fast MAR345 image plate detector with acquisition times between 0.5 and 15 min (typically 2-5 min). Electronic structure calculations were performed using the Wien2k code [31]; on basis of the density functional theory (DFT) and the local spin density approximation (LSDA) plus the multiorbital mean-field Hubbard model (LSDA + U, U > 2.5 eV) and the quantum chemical CRYSTAL code the *exact exchange* Becke three parameter Lee-Yang-Parr (B3LYP) hybrid functional [32]. For all phases a geometry optimization was performed and the phase stabilities were determined]. For all phases a geometry optimization was performed and the phase stabilities were determined. The LSDA + U approach allows the most important on-site correlations. To account for on-site correlation, the LDA + U method with U from 1 to 10 eV was used in the self-interaction correction scheme with J = 0. The number of k points were 12 * 12 *12, and the muffin tin radii were chosen for Fe = 2.5 Å and for Se = 2.24 Å. The structural data of the hexagonal phase are a = 3.66 Å c = 5.565 Å.

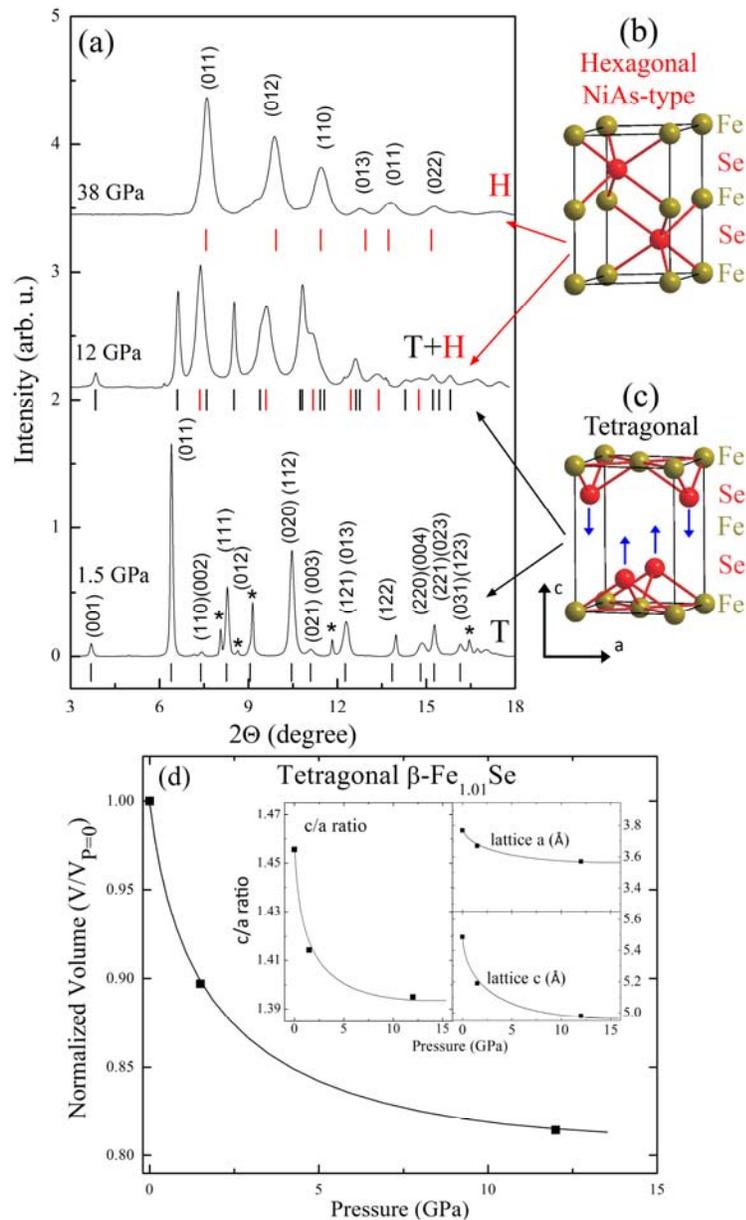

**Fig. 1**. (a) X-ray diffraction patterns of $Fe_{1.01}Se$ at various pressures. At 1.5 GPa, only the tetragonal form is present [shown in panel (c)]. At 12 GPa, the sample contains a mixture of tetragonal and hexagonal [NiAs-type, shown in panel (b)] forms. By 38 GPa the sample is solely hexagonal. The tick marks below the patterns show the calculated reflection positions for tetragonal (black), and hexagonal (red) phases. Reflections denoted with asterisks on the 1.5 GPa pattern originate from the Re-gasket. (d) Tetragonal $Fe_{1.01}Se$ undergoes a nearly 20% volume reduction under pressure, accompanied by a large decrease in the c/a ratio (inset), indicative of a collapse of the van-der-Waals-bonded interlayer region (blue arrows, panel (c)).

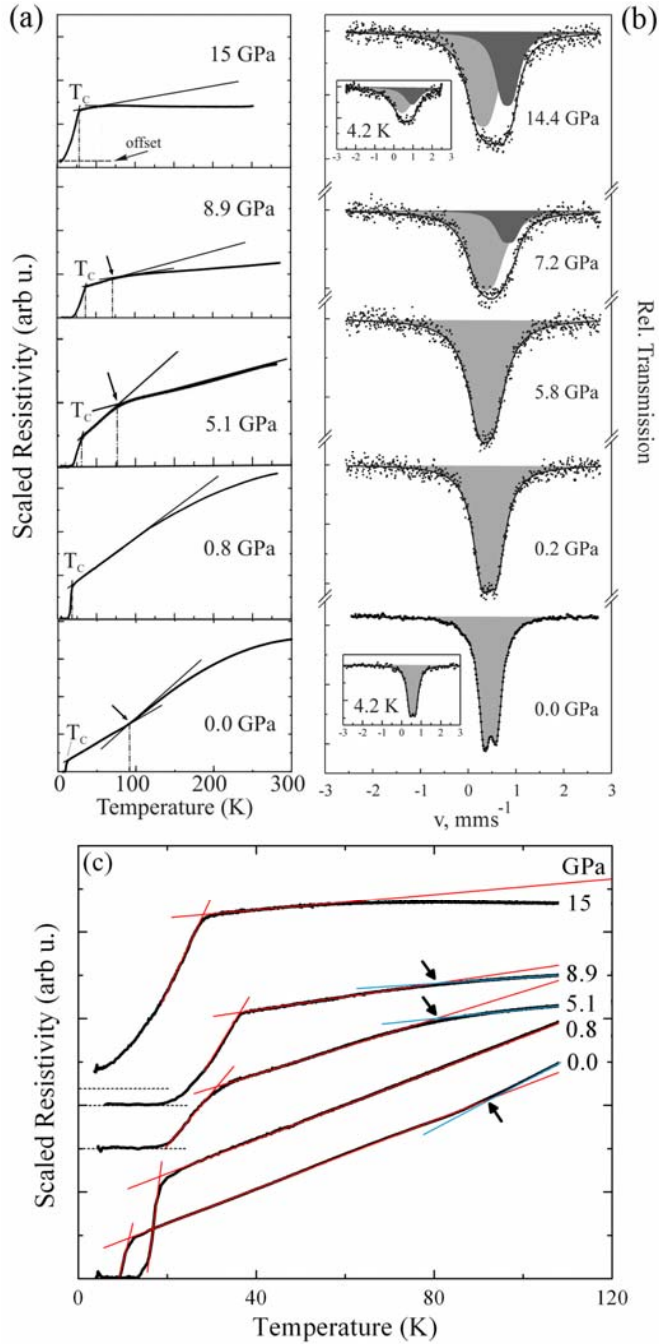

**Fig. 2** (a) Temperature dependent resistance curves of $Fe_{1.01}Se$ measured at different pressures. There is a systematic change in $T_c$, as well as the onset temperature of the low temperature linear resistivity regime (indicated by arrows). (b) $^{57}Fe$ Mössbauer spectra at $295K$ under various pressures (data for $4.2\ K$ shown in insets). Under no conditions is a sextet, indicative of static magnetic order, observed. The dark gray shading indicates the signal arising from the appearance of the high pressure hexagonal phase. (c) Detailed views of the superconducting transition at various pressures. Arrows indicate the onset temperatures of the linear temperature dependence of the resistivity. There is a change in concavity in the transition to linear behavior as the applied pressure is increased. Some curves have been offset for clarity and dotted lines equal zero in those cases.

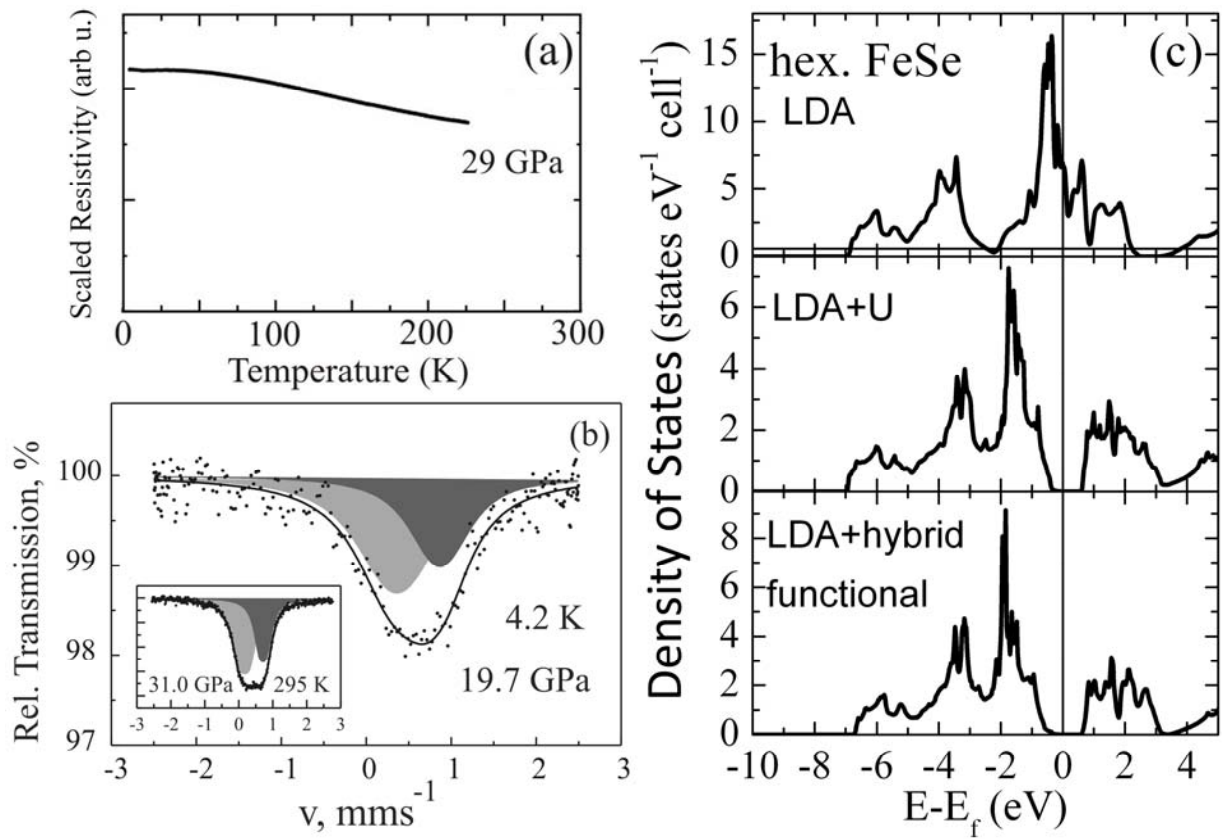

**Fig. 3** (a) An increasing phase fraction of the hexagonal form of iron selenide results in the onset semiconducting behavior by 30 GPa. (b) The Mössbauer spectrum of the sample at 19.7 GPa and 4.2 K shows two paramagnetic lines corresponding to the tetragonal (light gray) and hexagonal (dark gray) forms (spectrum at 295 K and 31.0 GPa shown in the inset). These data imply that the NiAs-type hexagonal phase is semiconducting without long range magnetic order. (c) This behavior cannot be explained under the local density approximation (LDA), which predicts metallic behavior due to the non-zero density of states at the Fermi level. The inclusion of electron correlations, either by LDA+U (U ~ 4 eV) or exact exchange (B3LYP functional), is necessary to reproduce the observed phenomenology.

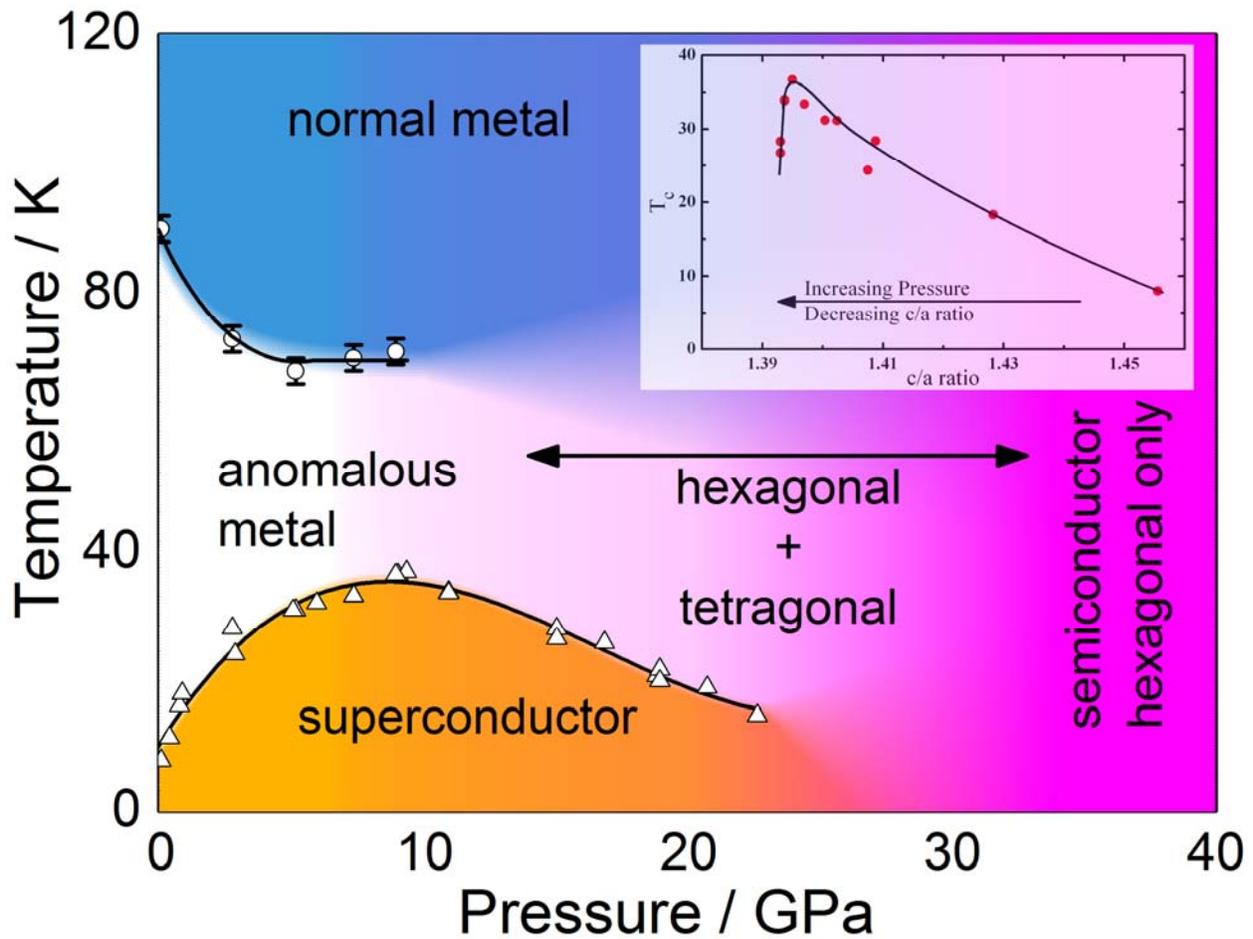

**Fig. 4** Electronic phase diagram of $Fe_{1.01}Se$ as a function of pressure. At low temperatures there is an anomalous metal regime where the resistivity is linear in temperature (see Fig. 2) prior to the onset of superconductivity. The maximum $T_c$ observed is 36 K at 8.9 GPa. At high pressures the sample is solely hexagonal, and shows semiconducting behavior. The inset shows the strong dependence of $T_c$ on the crystallographic c/a ratio.